\documentclass[conference]{IEEEtran}

\usepackage{hyperref}
\usepackage{url}

\usepackage{graphicx}
\usepackage{caption}
\usepackage{subcaption}
\usepackage{wrapfig}
\usepackage[dvipsnames]{xcolor}
\usepackage{amsmath,amssymb,amsfonts}
\usepackage{algorithm}
\usepackage{makecell}
\usepackage{algpseudocode}
\usepackage{tikz}
\usepackage{pgfplots}
\usepackage{calc}
\usepackage{balance} 
\usepackage{enumitem} 
\usepackage{amsmath}
\usepackage{xcolor}
\usepackage{booktabs}
\usepackage{multirow}
\usepackage{amsthm} 
\usepackage[capitalize,noabbrev]{cleveref}
\usepackage{soul}

\definecolor{codegreen}{rgb}{0,0.6,0}
\definecolor{codegray}{rgb}{0.5,0.5,0.5}
\definecolor{codepurple}{rgb}{0.58,0,0.82}
\definecolor{backcolour}{rgb}{0.95,0.95,0.92}
\definecolor{jcred}{HTML}{e31a1c}
\definecolor{jcgreen}{HTML}{33a02c}
\definecolor{jcblue}{HTML}{1f78b4}
\definecolor{jcorange}{HTML}{ff7f00}
\definecolor{jcpurple}{HTML}{6a3d9a}
\definecolor{jcbrown}{HTML}{b15928}
\definecolor{jclightred}{HTML}{fb8072}
\definecolor{jclightgreen}{HTML}{b3de69}
\definecolor{jclightblue}{HTML}{80b1d3}
\definecolor{jclightorange}{HTML}{fdb462}
\definecolor{jclightpurple}{HTML}{bebada}
\definecolor{jcredl}{HTML}{fb8072}
\definecolor{jcgreenl}{HTML}{b3de69}
\definecolor{jcbluel}{HTML}{80b1d3}
\definecolor{jcbluell}{HTML}{ece7f2}
\definecolor{jcorangel}{HTML}{fdb462}
\definecolor{jcpurplel}{HTML}{bebada}
\definecolor{jcbluem}{HTML}{488bb8}
\definecolor{jcyellow}{HTML}{ffff99}

\usepackage{pifont}
\newcommand{\cmark}{\textcolor{jcgreen}{\ding{51}}}%
\newcommand{\xmark}{\textcolor{jcred}{\ding{55}}}%

\newcommand*\best[1]{\textcolor{jcgreen}{\bf #1}}

\newcommand{\quant}[1]{\textcolor{jcblue}{#1}}
\newcommand{\apprx}[1]{\textcolor{jcgreen}{\bf #1}}

\begin{document}

\title{Refining Datapath for Microscaling ViTs}
\author{
\IEEEauthorblockN{Can Xiao}
\IEEEauthorblockA{
\textit{Imperial College London} \\
cx922@ic.ac.uk}
\and
\IEEEauthorblockN{Jianyi Cheng}
\IEEEauthorblockA{
\textit{University of Edinburgh} \\
jianyi.cheng@ed.ac.uk}
\and
\IEEEauthorblockN{Yiren Zhao}
\IEEEauthorblockA{
\textit{Imperial College London} \\
a.zhao@imperial.ac.uk}
}

\maketitle

\begin{abstract}
Vision Transformers (ViTs) leverage the transformer architecture to effectively capture global context, demonstrating strong performance in computer vision tasks.
A major challenge in ViT hardware acceleration is that the model family contains complex arithmetic operations that are sensitive to model accuracy, such as the Softmax and LayerNorm operations, which cannot be mapped onto efficient hardware with low precision. 
Existing methods only exploit parallelism in the matrix multiplication operations of the model on hardware and keep these complex operations on the CPU.
This results in suboptimal performance due to the communication overhead between the CPU and accelerator.
Can new data formats solve this problem?

In this work, we present the first open-source ViT accelerator that maps all operations of the ViT models onto FPGAs.
We exploit a new arithmetic format named Microscaling Integer (MXInt) for datapath designs and evaluate how different design choices can be made to trade off accuracy, hardware performance, and hardware utilization.
Our contributions are twofold.
First, we quantize ViTs using the MXInt format, achieving both high area efficiency and accuracy.
Second, we propose MXInt-specific hardware optimization that map these complex arithmetic operations into custom hardware.
Within 1\% accuracy loss, our method achieves at least 93$\times$ speedup compared to Float16 and at least 1.9$\times$ speedup compared to related work.
\end{abstract}

\IEEEpeerreviewmaketitle

\section{Introduction}
\label{sec:introduction}

Hardware acceleration for transformers has shown significant performance benefits compared to general processors~\cite{zeng2024flightllm, dong2023heatvit, li2022auto}, among which Vision Transformers (ViTs) offer promising performance for capturing global image relationships \cite{alexey2020image}. Compared to traditional Convolutional Neural Networks (CNNs), ViTs present new model features: 1) these models often contain millions of parameters, leading to a large memory size; and 2) they contain non-linear operations, requiring complex hardware operator designs.

Traditional techniques for ViT acceleration focus on {\em 1) integer quantization} and {\em 2) datapath optimization}, exploiting the approximation tolerance of ViT models. 
First, integer quantization represents numbers as small integers, optionally with a scaling factor, leading to both smaller memory and circuit area~\cite{dong2023heatvit, li2022auto}.
Second, datapath optimization determines new designs with simpler logic and similar results, leading to a smaller circuit area~\cite{yang2024sda}. 

Still, non-linear operations in ViT, such as LayerNorm and Softmax, face challenges in efficient acceleration.
These operations contain complex mathematical operations, such as {\em exp()} and {\em sqrt()}, and require large value ranges, restricting existing integer quantization. 
Existing design methods rely on the CPU and only accelerate part of the ViT models in FPGA fabric~\cite{dong2023heatvit, li2022auto}.
This leads to a working but complex system design with suboptimal performance due to the communication overhead between the CPU and the accelerator.

In this work, we unlock this by exploiting a recently studied data format named Microscaling Integers (MXInt)~\cite{darvish2020pushing}. 
The MXInt format shares an exponent among a block of integer values, forming a more compact floating-point format.
This reduces memory size while maintaining high model accuracy~\cite{zhang2023revisiting}.
The existing work in MXInt hardware mapping focuses on matrix multiplications~\cite{samson2024exploring}, but it remains an open question regarding optimizing non-linear operations in MXInt. 

\begin{table}[]
\caption{Our MXInt design method maps all non-linear operations in ViTs into efficient hardware, achieving lower bitwidths than traditional fixed-point designs.}
\label{tab:related_work}
\resizebox{0.48\textwidth}{!}{%
\begin{tabular}{lccccccc}
\toprule
\multirow{2}{*}{Methods} & \multirow{2}{*}{\begin{tabular}[c]{@{}c@{}}Precision\\ on fabric\end{tabular}} & \multicolumn{2}{c}{LayerNorm} & \multicolumn{2}{c}{GELU} & \multicolumn{2}{c}{Softmax} \\
\cmidrule(lr){3-4}
\cmidrule(lr){5-6}
\cmidrule(lr){7-8}
 &  & Fabric & Bitwidth & Fabric & Bitwidth & Fabric & Bitwidth \\
\midrule
AutoViTAcc~\cite{li2022auto} & fixed-point & \xmark & 8 & \xmark & 8 & \xmark & 8 \\
\cite{huang2023integer} & fixed-point & \cmark & 8 & \cmark & 8 & \cmark & 8 \\
HeatViT~\cite{dong2023heatvit} & fixed-point & \xmark & 8 & \cmark & 8 & \cmark & 8 \\
SDA ~\cite{yang2024sda} & fixed-point & \cmark & 8 & \cmark & 8 & \cmark & 8 \\
\midrule
Our Work & {\bf \color{jcgreen} MXInt} & \cmark & 
{\bf \color{jcgreen} 5} 
& \cmark & 
{\bf \color{jcgreen} 5} 
& \cmark & 
{\bf \color{jcgreen} 2} 
\\
\bottomrule
\end{tabular}
}
\end{table}
\begin{figure*}
    \begin{subfigure}[b]{0.3\textwidth}
        \begin{subfigure}[b]{\textwidth}
    \centering
    \includegraphics[width=0.7\textwidth]{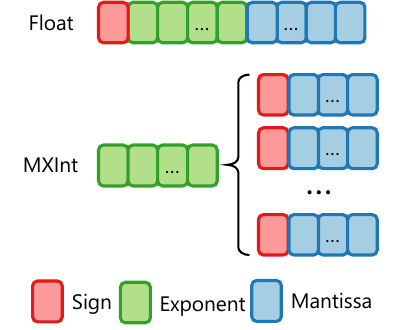}
    \caption{A standard floating point format and a MXInt format.}
    \label{fig:format}
    \end{subfigure}
    \begin{subfigure}[b]{\textwidth}
    \centering
\resizebox{\textwidth}{!}{%
    \begin{tabular}{ccrr}
\toprule
Formats & Config & \multicolumn{1}{c}{Accuracy} & \makecell[c]{Memory \\ Density} \\
\midrule
{\tt FP32} & - & \best{81.80\%} & 1$\times$ \\
\tt{Int16} & W16A16 & 80.05\% & 2$\times$ \\
\midrule
\tt{MXInt8} & W6.03/A8.5 & \best{81.72\%} & 4.99$\times$ \\
\bottomrule
\end{tabular}
}
    \caption{Quantization results of DeiT Base~\cite{touvron2022deit} on ImageNet~\cite{deng2009imagenet}. MXInt balances model accuracy and memory efficiency, leading to a better choice of formats for ViT quantization.
    The decimal part derived from the shared exponent value.}
    
    \label{tab:mxint_accuracy}
\end{subfigure}
    \end{subfigure}
    \begin{subfigure}[b]{0.01\textwidth}
    ~
    \end{subfigure}
    \begin{subfigure}[b]{0.3\textwidth}
    \begin{algorithmic}[1] \footnotesize
    \Require $X$ \Comment{Input Features}
    \Require $H$ \Comment{Number of heads}
    \Require $L$ \Comment{Number of hidden layers}
    \State $\quant{X_n} \gets \apprx{LayerNorm}(\quant{X}) $
    \For{$i \in [0, H)$}
    \State $\quant{Q_i} \gets \quant{W_{Q_i}} \cdot \quant{X_n}$
    \State $\quant{K_i} \gets \quant{W_{K_i}} \cdot \quant{X_n}$ 
    \State $\quant{V_i} \gets \quant{W_{V_i}} \cdot \quant{X_n}$ 
    \State $\quant{A_i} \gets \frac{\quant{Q_i} \cdot \quant{K_i}^T}{\sqrt{d_k}} $
    \State $\quant{\hat{A}_i} \gets \apprx{Softmax}(\quant{A_i}) $
    \State $\quant{B_i} \gets \quant{\hat{A}_i}\cdot \quant{V_i}$
    \EndFor
    \State $\quant{B_c} \gets Concat(\quant{B_0}.. \quant{B_{H-1}}) $
    \State $\quant{B_o} \gets \quant{W_0} \cdot \quant{B_c}$
    \State $\quant{B_n} \gets \apprx{LayerNorm}(\quant{B_o} + \quant{X_n}) $
    \State $\quant{U} \gets \quant{W_U} \cdot \quant{B_n}$
    \State $\quant{D} \gets \quant{W_D} \cdot (\apprx{GELU}(\quant{U}))$
    \State $\quant{O} \gets \quant{D} + \quant{B_n}$
    \State \Return $\quant{O}$
    \end{algorithmic}
    \caption{An algorithm view of a block in the DeiT model.
    Values highlighted in \quant{\em blue} represent quantized values, and operations highlighted in \apprx{green} represent mxint specific approximated operations.}
    \label{fig:motivation_alg}
    \end{subfigure}
    \begin{subfigure}[b]{0.01\textwidth}
    ~
    \end{subfigure}
    \begin{subfigure}[b]{0.35\textwidth}
    \includegraphics[width=0.9\textwidth]{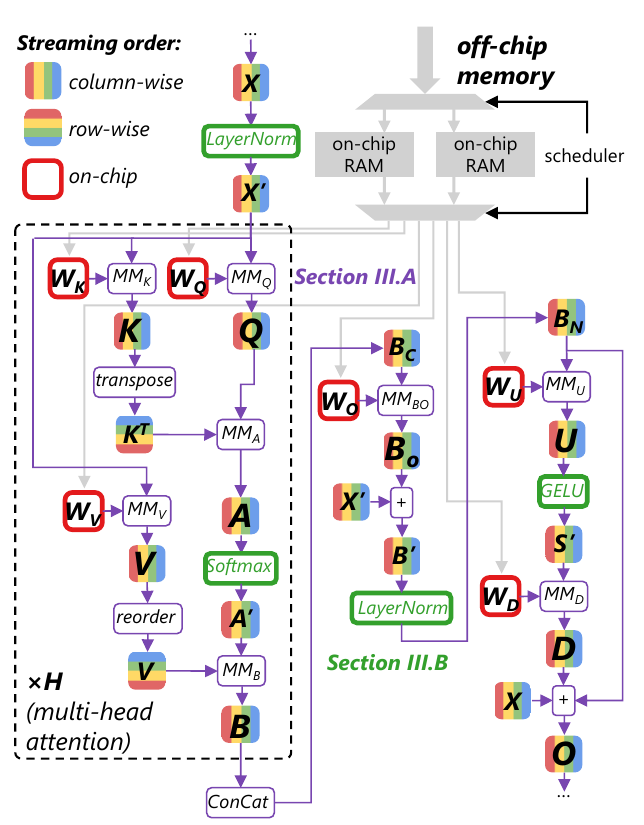}
    \caption{An architecture view of the proposed hardware accelerator.
    The proposed architecture pipelines the model in a hierarchical dataflow, and tailors each operation for high area efficiency.}
    \label{fig:motivation_hw}
    \end{subfigure}
    \caption{Motivating example: dataflow hardware acceleration of a ViT in MXInt.}
    \label{fig:motivation}
    \end{figure*}
    
In this work, we present fully quantized ViTs in MXInt and propose datapath optimization techniques for the efficient acceleration of both linear and non-linear operations.
To ensure fairness in quantization, we focus on post-training quantization (PTQ) and compare accuracy without fine-tuning, as fine-tuning entails complex training techniques that may vary from model to model.
Following prior work on MXInt quantization~\cite{zhang2023revisiting}, we restrict the accuracy loss of the final design to within 1\% to preserve high model accuracy.
We then explore datapath optimization opportunities and show how to efficiently map MXInt operators into efficient hardware, including non-linear operations. 
This leads to fully hardware-accelerated ViTs, meaning that all operations are mapped into efficient hardware, as illustrated in Table~\ref{tab:related_work}. Our contributions are as follows:
\begin{itemize}
    \item We propose MXInt-based ViT accelerators, reducing the memory size up to 4.99$\times$ within 1\% accuracy loss;
    \item We propose MXInt-specific datapath optimization for accuracy-sensitive arithmetic operators in the ViTs (e.g. GELU, Softmax and LayerNorm), reducing the area at least by 16$\times$ within 1\% accuracy loss; and
    \item Over a set of models, we show our design methods achieve at least 93$\times$ speedup compared to Float16 and at least 1.9$\times$ speedup compared to related work.
\end{itemize}
The rest of the paper is organized as follows. 
Section~\ref{sec:background} illustrates an overview of the proposed accelerator architecture.
Section~\ref{sec:method} describes our hardware architecture and design choices in detail.
Section~\ref{sec:experiments} evaluates the effectiveness of our work and compares performance with related work.
Section~\ref{sec:related_work} compares our work with related work qualitatively.












    
    

\section{Motivation}
\label{sec:background}

{\em Why MXInt?}
Here we first introduce the MXInt format. 
\Cref{fig:format} compares the standard floating-point numerical format with the MXInt format.
The top of the figure illustrates the standard floating-point format, which contains a sign bit, an exponent, and a mantissa~\cite{kahan1996ieee}.
The MXInt format has similar components to the standard floating-point format but allows a block of values to share the exponent, where the block size is user-defined. 
MXInt provides finer granularity compared to the traditional fixed-point format that shares a scaling factor across all tensor values. 
Prior work~\cite{zhang2023revisiting} shows that MXInt can achieve both high model accuracy and high memory density for software LLM quantization.
Our observation from Figure~\ref{tab:mxint_accuracy} verifies that this still holds for ViT models.

{\em How to map into MXInt?}
\Cref{fig:motivation} provides a high-level overview of the proposed accelerator architecture design.
\Cref{fig:motivation_alg} illustrates an algorithmic view of a transformer block in ViT models.
There are various design choices on hardware architecture for the ViT accelerators, such as systolic arrays and custom dataflow architectures.
The benefits of dataflow architectures have been widely studied in the literature~\cite{chen2024understanding}.
Here we choose the dataflow architecture due to its high throughput and low control flow overhead.

The corresponding accelerator architecture is shown in \Cref{fig:motivation_hw}.
In the figure, each operation in \Cref{fig:motivation_alg} is mapped into a hardware operator unit with a dataflow interface.
Each hardware unit is dedicated to computing a specific function for its inputs, minimizing control flow overhead.
In our design, the parameters are initially stored off-chip due to the large size required by the ViT models.
A predefined scheduler is implemented to prefetch the parameters to on-chip memory through a ping-pong buffer, as shown at the top right of \Cref{fig:motivation_hw}.
Similar to most dataflow architectures, the tensor is tiled due to its large size and streamed into the accelerators for deeply pipelined computation.
Design choices for tiling sizes and streaming order have been widely studied in prior work~\cite{venieris2016fpgaconvnet, umuroglu2017finn, ye2023hida}.
These are outside our scope, while we focus on the datapath optimization of these operators.
We will describe our MXInt quantization and linear operator implementation in Section~\ref{sec:mxint_quantization}.

\begin{figure*}
\centering
\begin{subfigure}[b]{0.5\textwidth}
\centering
\includegraphics[scale=0.4]{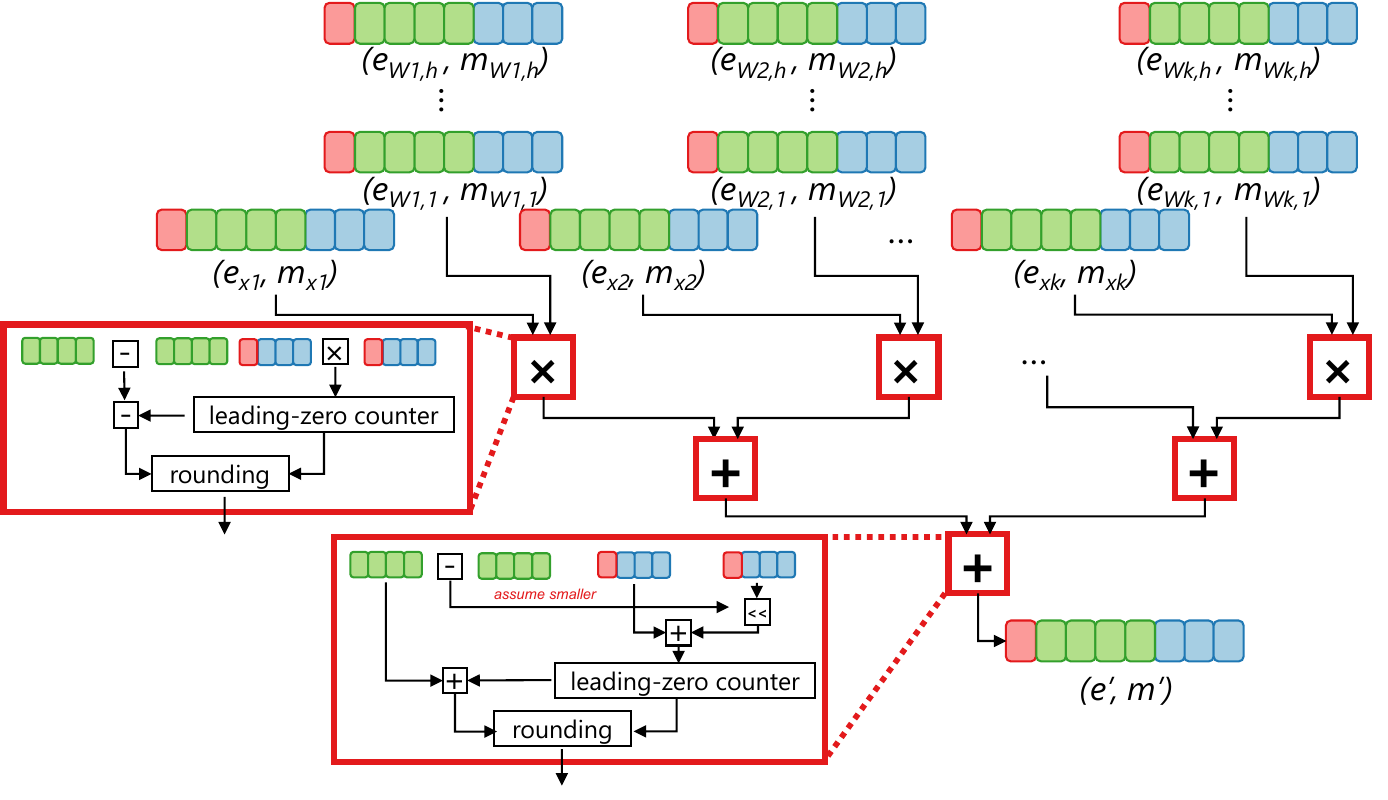}
\caption{Dot product in Float relies on area-expensive operators.}
\label{fig:linear_float}
\end{subfigure}
\hfill
\begin{subfigure}[b]{0.45\textwidth}
\centering
\includegraphics[scale=0.44]{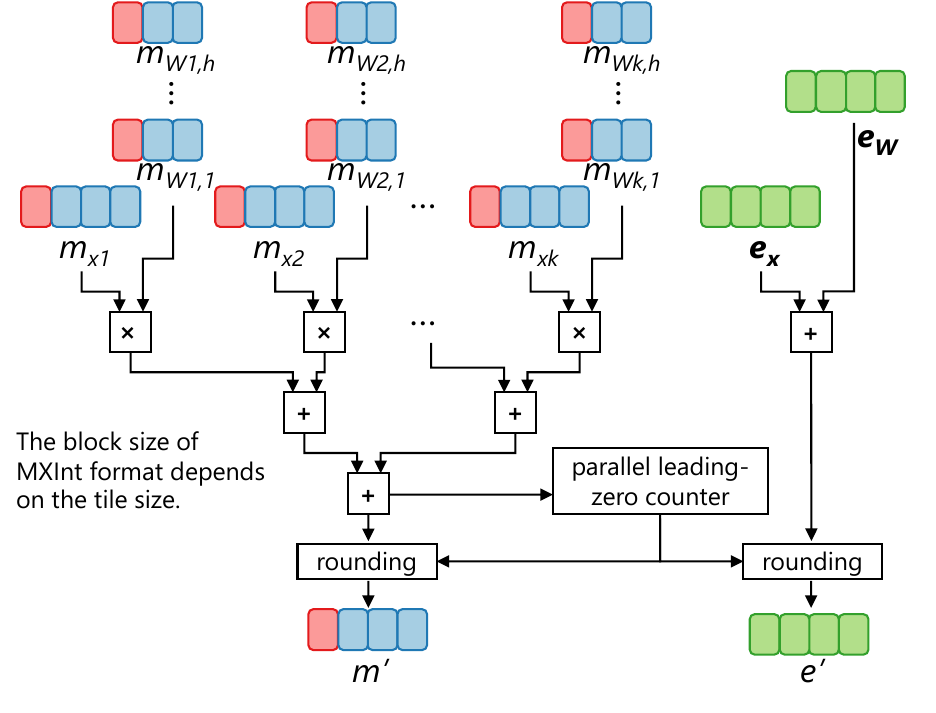}
\caption{Dot product in MXInt contains integer-only operators.}
\label{fig:linear_mxint}
\end{subfigure}
\caption{Comparison between the standard floating-point format and MXInt on the dot product operator.}
\label{fig:dot_product}
\end{figure*}

{\em How to optimize MXInt hardware?}
The ViTs also contain non-linear operations, and the precision required by these operations is often high due to their high sensitivity to model accuracy.
This leads to significant circuit area due to both high precision and arithmetic complexity.
We will show how to lift this restriction by optimizing the datapath of non-linear operators in Section~\ref{sec:hardware_approximation}.

\section{Methodology}
\label{sec:method}


\subsection{Opportunities and Challenges for MXInt}
\label{sec:mxint_quantization}

The hardware design of efficient MXInt operators faces several optimization opportunities regarding the block-sharing feature of the MXInt data format. 
Traditional operators compute on individual values independently because their values are presented in standalone formats. 
Such a design method is inefficient for MXInt values because the common results of the shared exponent could be shared across the values in the same block.

For simplicity, here we show a dot product design example to illustrate how mapping into MXInt causes hardware design improvements.
\Cref{fig:linear_float} shows a traditional implementation of a dot product unit with $k$ values computing in parallel.
The precision of the values is in standard floating-point to preserve high model accuracy.
In the data path, standard floating-point multipliers and adders are used for computation.
Using these operators often leads to significant area overhead because the normalization between values to the same dynamic range requires complex dynamic shift logic~\cite{coward2023automating}.
The normalization circuit is required at both the input and the output of each operator, leading to huge area overhead in total. 
There has been effort in optimizing normalization for individual floating-point operators~\cite{alexandridis2024online}, but they miss MXInt-specific optimizations such as exponent sharing.

We design fully customized operators for MXInt values, as illustrated in \Cref{fig:linear_mxint}.
Here, we assume that the exponent of $x$ is shared by $k$ values, and the exponent of $W$ is shared by $h \times k$ values.
In the dot product unit, all the operators perform either fixed-point or logical operations.
The overhead caused by the expensive dynamic shift logic in standard floating-point is reduced from two aspects: 1) only one dynamic shift operator is required per block, where its result is shared within the block; and 2) the bitwidth of the mantissas in MXInt is often small after quantization, making the transformation into unrolled constant shifts affordable. 
This enables us to compute values with dynamic ranges in efficient hardware while preserving high accuracy.

A key challenge is to balance the precision and circuit area by determining an efficient mantissa bitwidth for MXInt values. 
A small bitwidth in MXInt mantissas could lead to quantization errors, while a large bitwidth could lead to a large operator area.
Particularly, the bitwidth required by the accumulator used in the linear operations scales with the tensor sizes.
A significantly large tensor with a large variance in element values could cause the accumulator to both underflow and overflow.

In this work, we focus on the hardware datapath optimization of non-linear operations, as linear operations have already been studied in related work~\cite{samson2024exploring}.
For linear operations, we rely on greedy search in software quantization to determine the minimal bitwidth of the mantissa to preserve high accuracy within a 1\% loss. 
We observed that the shared exponent can effectively handle overflow with negligible effects on model accuracy, but the underflow requires more bits.
Therefore, we empirically determine the additional bits required by the accumulator operator and expand the bitwidth of the accumulator to minimize quantization loss.
Particularly for DeiT models, we set the mantissa of the accumulator in all the linear operators to be 12 bits to perform lossless addition.
A similar method is applied to other compute-intensive operators, such as convolution operators.

Finally, different shared block sizes in MXInt between layers could lead to additional hardware logic to ungroup and regroup these values between computations.
This is more related to the control path of the circuit and is out of the scope of our work.
In this work, we choose the block size to always be the same as the tile size, as shown in \Cref{fig:linear_mxint} to minimize the control flow overhead on grouping values.
For instance, a linear operator has exponents shared among 16 and 256 values in activation and weights, respectively. In the rest of the section, we will describe our optimization techniques for the datapath of non-linear operators.

\subsection{Datapath Optimization in MXInt}
\label{sec:hardware_approximation}


In this section, we now show how to optimize all these non-linear functions in the mantissa domain with fewer numbers of bits.
Our work focuses on optimizations for ViTs, but it could also be applied to MXInt accelerators for other ML models, leading to different results of bitwidth.
Here we take three core operators from three representative classes of operations for illustration: LayerNorm, Softmax, and GELU. These operations are commonly seen in ViT models and require complex arithmetic circuits to compute.

\subsubsection{LayerNorm Approximation}

LayerNorm operations have been widely used in transformers, including ViTs, which scale values in a tensor to a fixed range~\cite{layernorm}. The standard expression of the LayerNorm operation is presented as follows.
\begin{align}
    y = \frac{x - E(x)}{\sqrt{Var(x)+\epsilon}} \gamma + \beta  \label{eqn:layernorm}
\end{align}
$x$ and $y$ denote the input and output tensors, and $E(.)$ computes the expectation and $Var(.)$ computes the variance, respectively. $\epsilon$, $\gamma$, and $\beta$ are constants in the model.

\begin{figure}[t]
    \centering
    \includegraphics[width=0.9\linewidth]{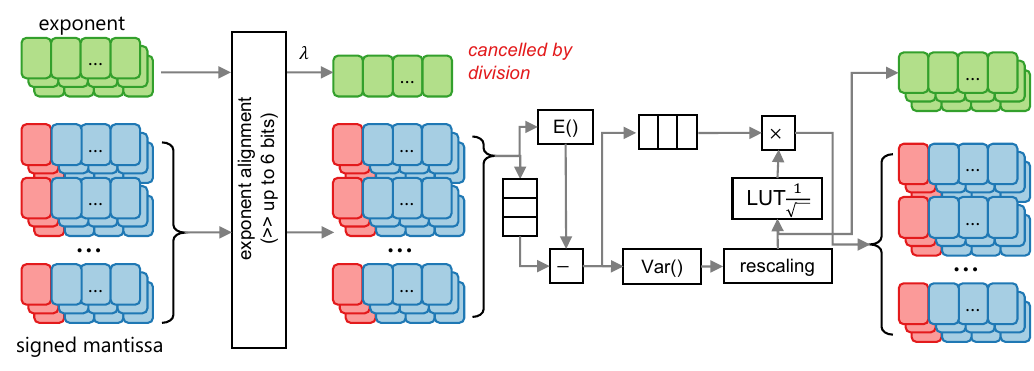}
    \caption{Optimized datapath for MXInt LayerNorm.}
    \label{fig:layernorm}
\end{figure}

Existing approaches, including \cite{huang2023integer} and SDA~\cite{yang2024sda}, focus on integer-based datapath optimization for hardware acceleration, but their bitwidth is still large compared to our MXInt designs. This is because the absence of dynamic range in integer format requires more bits to represent the same range of values. 

We now describe our optimization for LayerNorm in MXInt. 
Benefiting from the small mantissa bitwidth of MXInt, we separate the computation of the shared exponent and mantissas.
An MXInt value can be represented as follows.
\begin{align}
   x = 2^{x_e}x_m 
\end{align}
$x_e$ and $x_m$ are the unsigned shared exponent and signed mantissa for a value $x$, where $x_e$ is shared by a group of values. When computing Softmax, we are effectively dealing with MXInt values coming from different blocks that are scaled by a set of different exponent values $(x_e^0, x_e^1, ...)$. As shown in \Cref{fig:layernorm}, we apply a re-quantization step that forces these values from different groups to use the same exponent ($x^{max}_{e}$) by dynamically right-shifting on mantissa values. Effectively, for each individual MXInt value, we have:
\begin{align}
   x = 2^{x_e}x_m = 2^{x^{max}_{e}}x_{m'} = \lambda x_{m'}
\end{align}
where $x_{m'}$ is the right-shifted version of $x_m$, and the right-shifting amount depends on $x^{max}_{e} - x_e$. As shown in \Cref{fig:layernorm}, since $x^{max}_{e}$ is now shared across all values that are inputs to the Softmax function, we can effectively treat it as a constant $\lambda = x^{max}_{e}$ for the ease of expression.

We then substitute the expression of $x$ into Equation~\ref{eqn:layernorm} and extract the exponent as follows.
\begin{align}
    y =& \frac{\lambda x_m - E(\lambda x_m)}{\sqrt{Var(\lambda x_m)+\epsilon}} \gamma + \beta \\
    =& \frac{\lambda (x_m - E(x_m))}{\sqrt{Var(\lambda x_m)+\epsilon}} \gamma + \beta 
\end{align}
We approximate $\epsilon$ to be zero so that we can extract the exponent from the square root function.
\begin{align}
    \approx& \frac{\lambda (x_m - E(x_m))}{\lambda \sqrt{Var(x_m)}} \gamma + \beta \\
    \approx& \frac{(x_m - E(x_m))}{\sqrt{Var(x_m)}} \gamma + \beta
\end{align}
A key novelty of our work is that we convert an MXInt-based LayerNorm operation into an integer-only operator dealing with only the mantissa component ($x_m$) in the MXInt format.
This significantly simplifies the circuitry required to implement the LayerNorm function.

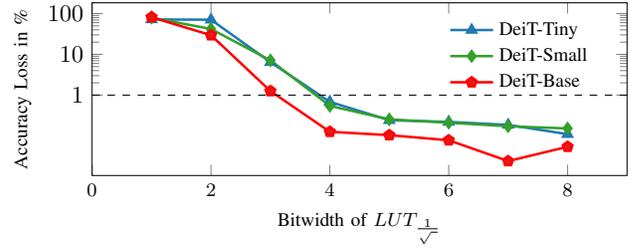
\begin{figure}
    \centering
\begin{tikzpicture}[thick,scale=0.95, every node/.style={scale=0.95}]
\pgfplotsset{every x tick label/.append style={font=\footnotesize}, compat=1.3}
\begin{semilogyaxis}[
    height=40mm,
    width=0.5\textwidth,
    xlabel={\footnotesize Bitwidth of $LUT_{\frac{1}{\sqrt{~}}}$},
    ylabel={\footnotesize Accuracy Loss in \%},
    xmin=0, xmax=9,
    ytick={1, 10, 100},
    ymode=log, log ticks with fixed point, 
    legend cell align={left},
    legend style={at={(0.95,0.95)},anchor=north east, draw=none, fill=none, font=\footnotesize},
    ]

\addplot[jcblue, draw=jcblue, line width=1pt, mark=triangle*] coordinates {
    (1, 72.044)
    (2, 70.86)
    (3, 06.446)
    (4, 00.684)
    (5, 00.242)
    (6, 00.22)
    (7, 00.186)
    (8, 00.11)
};

\addplot[jcgreen, draw=jcgreen, line width=1pt, mark=diamond*] coordinates {
    (1, 79.71)
    (2, 41.974)
    (3, 07.134)
    (4, 00.544)
    (5, 00.254)
    (6, 00.21)
    (7, 00.172)
    (8, 00.152)
};

\addplot[jcred, draw=red, line width=1pt, mark=pentagon*] coordinates {
    (1, 81.734)
    (2, 29.564)
    (3, 01.258)
    (4, 00.126)
    (5, 00.104)
    (6, 00.078)
    (7, 00.024)
    (8, 00.054)
};
\addplot [dashed, black] coordinates {
    (0.005, 1) (9, 1)
};

\legend{
    DeiT-Tiny, 
    DeiT-Small, 
    DeiT-Base
}
\end{semilogyaxis}
\end{tikzpicture}    
    \caption{Accuracy loss over different bitwidth of $LUT_{\frac{1}{\sqrt{~}}}$.}
    \label{fig:layer_norm_search}
\end{figure}

\begin{table}
\centering
\caption{Comparison with related work on LayerNorm optimization for DeiT-Tiny. Our work achieves the minimal accuracy loss on fabric at the smallest bitwidth.}
\label{tab:layernorm_comp}
\resizebox{\linewidth}{!}{%
    \begin{tabular}{lcrrr}
        \toprule
        Methods & On Fabric & \makecell[c]{Bitwidth} & \makecell[c]{Accuracy} & \makecell[c]{Accuracy Loss} \\
        \midrule
        Original & \xmark & 16 & 72.13\% & - \\
        Auto-ViT-Acc\cite{li2022auto} & \xmark & 16 & 72.13\% & 0 \\
        HeatViT~\cite{dong2023heatvit} & \xmark & 16 & 72.13\% & 0 \\
        \cite{huang2023integer} & \cmark & 8 & 71.66\% & 0.472\%\\
        SDA~\cite{yang2024sda} & \cmark & 8 & 71.66\% & 0.472\%\\
        \midrule
        Vanilla MXInt & \cmark& 13 & 72.078\% & 0.154\%\\
        Optimized MXInt & \cmark & {\bf \color{jcgreen} 5} & 71.89\% & {\bf \color{jcgreen} 0.242\%}\\
        \bottomrule
    \end{tabular}%
}
\label{tab:compar_sqrt}
\end{table}

Still, after computing the variance $Var(.)$, we end up with a large result in a larger bitwidth from the accumulator, determined by the tensor size. 
This is followed by a division and a square root operation ($\frac{1}{\sqrt{Var(.)}}$).
Existing methods use large bitwidth in fixed-point numbers~\cite{huang2023integer} or cast the values to floating-point numbers to preserve high precision in a small bitwidth. We combine the best of both approaches, casting the values to a small floating-point format followed by a LUT-based method to avoid high computational overhead.

Specifically, we rescale (\textit{rescaling} in \Cref{fig:layernorm}) the variance ($Var(x_m)$), which is represented in the mantissa domain to a floating-point number $x^v_{m'}2^{x^v_{e'}}$.
\begin{align}
\frac{1}{\sqrt{x^v}} = (x^v)^{-1/2} = (x^v_m \cdot 2^{x^v_e})^{-1/2} = 2^{-x^v_e/2} \cdot (x^v_m)^{-1/2}.
\end{align}
Here only $x^v_m$ needs complex operations $\frac{1}{\sqrt{~}}$ and $x^v_e/2$ can be handled by shift. 
We use a LUT to represent the function $\frac{1}{\sqrt{~}}$ to reduce the computation overhead. 
\begin{align}
x = 
\begin{cases} 
LUT_{\frac{1}{\sqrt{~}}}(x^v_m) 2^{(-\frac{x^v_e}{2})}, & x^v_e \bmod 2 = 0, \\
LUT_{\frac{1}{\sqrt{~}}}(\frac{x^v_m}{2}) 2^{(-\frac{x^v_e+1}{2})}, & x^v_e \bmod 2 \neq 0. \\
\end{cases}
\end{align}
A key benefit of our approach is that the required LUT entries are significantly smaller due to excluding $x^v_e/2$, and the area of the dynamic shift operator in rescaling is small due to the low precision of mantissas.

\Cref{fig:layer_norm_search} illustrates the accuracy loss of the model over different LUT entries. We observe that a minimum of 4 bits is required by $x^v_m$ to preserve the model accuracy within a 1\% loss. \cref{tab:layernorm_comp} compares our optimization with related work on accuracy loss at the system level. We observed that 5 bits are required for MXInt ViTs when combined with optimizations on other layers. Still, our optimization saves a significant number of LUT entries compared to the vanilla MXInt operator implementation. Compared to related work using integer-based datapath optimization, our work achieves minimal accuracy loss with a significantly smaller bitwidth, thanks to the dynamic ranges provided by the shared exponent.


\subsubsection{GELU Approximation}

The mathematical definition of the GELU function involves complex arithmetic operations with high quantization sensitivity, particularly for small input values~\cite{gelu}. The standard formulation of GELU is as follows.
\begin{align}
\mathrm{GELU}(x)=\frac{x}{\sqrt{2\pi}} \int_{-\infty}^{x} e^{-t^2/2} dt
\end{align}
Existing work approximates it into a polynomial-based error function (erf) to reduce the computational overhead~\cite{dong2023heatvit, huang2023integer}.
\begin{align}
\mathrm{GELU}(x) \approx  \frac{x}{2} \left[ 1 + L_{erf}\left( \frac{x}{\sqrt{2}} \right) \right] \label{eqn:gelu_approx}
\end{align}
Benefiting from the small bitwidth of mantissas in MXInt, we further push the boundaries of such an approximation and propose a LUT-based method to map GELU into efficient hardware operators in MXInt. 

\begin{figure}[t]
    \centering
    \includegraphics[width=0.65\linewidth]{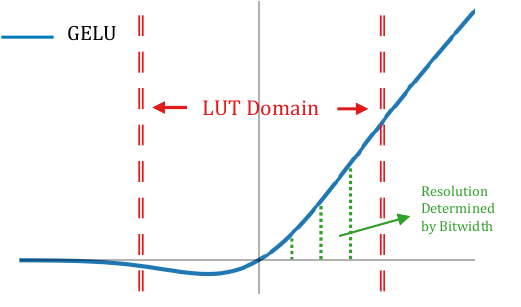}
    \caption{The LUT domain covers the non-linear region of the GELU function and its bitwidth affects the precision.}
    \label{fig:gelu_dis}
\end{figure}

\begin{figure}[t]
    \centering
    \includegraphics[width=0.8\linewidth]{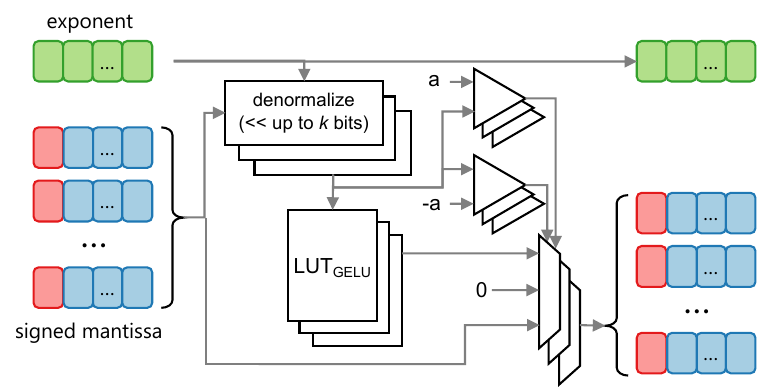}
    \caption{Optimized datapath for MXInt GELU. $k$ = LUT bitwidth + $log_2$(LUT domain) - 1.}
    \label{fig:gelu}
\end{figure}

A key challenge in the LUT-based optimization is that traditional designs still require input values containing both the exponent and the mantissa, leading to a significant number of entries.
However, this overhead may be affordable for MXInt designs, thanks to the small mantissa bitwidth of MXInt values.    
We present an efficient LUT-based optimization that partitions its input domain into three parts, as illustrated in \Cref{fig:gelu_dis}. The absolutely higher ranges on both ends are approximated to the ReLU function, and the small value range is mapped into a LUT.
\begin{align}
y = 
\left\{
\begin{array}{ll}
x, & x \geq a \\
LUT_{GELU}(x) & \text{if } -a < x < a \\
0, & x \leq a
\end{array}
\right.
 \end{align}

The optimized datapath is illustrated in \Cref{fig:gelu}. Since there is only a small difference between the input and the output of the GELU function, the exponent value does not change and is directly forwarded to the output. For the small values, each MXInt value as a floating-point number is cast to a fixed-point number and passes through the LUT.
The proposed design involves two design constraints.
First, $a$ defines the LUT domain that covers the non-linear range for small values, and the bitwidth of the LUT determines the resolution of the curve. Second, the bitwidth of the LUT determines the resolution of the curve.

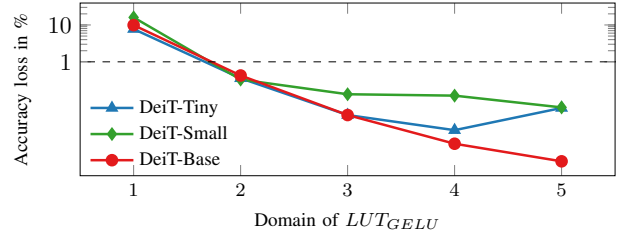
\begin{figure}
    \centering
\begin{tikzpicture}[thick,scale=0.95, every node/.style={scale=0.95}]
\pgfplotsset{every x tick label/.append style={font=\footnotesize}, compat=1.3}
\begin{semilogyaxis}[
    height=40mm,
    width=0.5\textwidth,
    xlabel={\footnotesize Domain of $LUT_{GELU}$},
    ylabel={\footnotesize Accuracy loss in \%},
    xmin=0.5, xmax=5.5,
    ytick={1, 10, 100},
    ymode=log, log ticks with fixed point, 
    legend cell align={left},
    legend style={at={(0.3,0.5)},anchor=north east, draw=none, fill=none, font=\footnotesize},
    ]

\addplot[jcblue, draw=jcblue, line width=1pt, mark=triangle*] coordinates {
    (1, 7.752)
    (2, 0.358)
    (3, 0.036)
    (4, 0.014)
    (5, 0.056)
};

\addplot[jcgreen, draw=jcgreen, line width=1pt, mark=diamond*] coordinates {
    (1, 16.018)
    (2, 0.331)
    (3, 0.132)
    (4, 0.121)
    (5, 0.058)
};
\addplot[jcred, draw=jcred, line width=1pt, mark=*] coordinates {
    (1, 9.892)
    (2, 0.426)
    (3, 0.036)
    (4, 0.006)
    (5, 0.002)
};


\addplot [dashed, black] coordinates {
    (0.005, 1) (9, 1)
};

\legend{
    DeiT-Tiny,
    DeiT-Small,
    DeiT-Base,
}

\end{semilogyaxis}
\end{tikzpicture}    
    \caption{Accuracy loss over different LUT domains for GELU. Bitwidth = 8.}
    \label{fig:gelu_clipping_search}
\end{figure}

\begin{figure}[t]
    \centering
\begin{tikzpicture}[thick,scale=0.95, every node/.style={scale=0.95}]
\pgfplotsset{every x tick label/.append style={font=\footnotesize}, compat=1.3}
\begin{semilogyaxis}[
    height=40mm,
    width=0.5\textwidth,
    xlabel={\footnotesize Bitwidth of $LUT_{GELU}$},
    ylabel={\footnotesize Accuracy loss in \%},
    ytick={1, 10, 100},
    xmin=0, xmax=9,
    ymode=log, log ticks with fixed point, 
    legend cell align={left},
    legend style={at={(0.95,0.95)},anchor=north east, draw=none, fill=white, font=\footnotesize},
    ]

\addplot[jcblue, draw=jcblue, line width=1pt, mark=triangle*] coordinates {
    (1, 71.17)
    (2, 48.594)
    (3, 04.922)
    (4, 00.888)
    (5, 00.132)
    (6, 00.012)
    (7, 00.024)
    (8, 00.032)
};

\addplot[jcgreen, draw=jcgreen, line width=1pt, mark=diamond*] coordinates {
    (1, 76.69)
    (2, 23.74)
    (3, 01.628)
    (4, 00.214)
    (5, 00.012)
    (6, 00.024)
    (7, 00.012)
    (8, 00.006)
};

\addplot[jcred, draw=red, line width=1pt, mark=pentagon*] coordinates {
    (1, 81.418)
    (2, 15.338)
    (3, 00.684)
    (4, 00.172)
    (5, 00.014)
    (6, 00.044)
    (7, 00.028)
    (8, 00.028)
};
\addplot [dashed, black] coordinates {
    (0.005, 1) (9, 1)
};

\legend{
    DeiT-Tiny, 
    DeiT-Small, 
    DeiT-Base
}
\end{semilogyaxis}
\end{tikzpicture}    
    \caption{Accuracy loss over different bitwidths for GELU. LUT domain = 3.}
    \label{fig:gelu_search}
\end{figure}
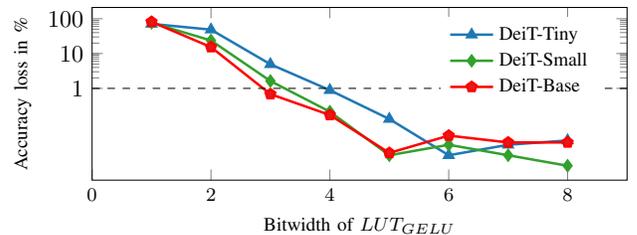

The design space explorations of both the LUT domain and its bitwidth are shown in \Cref{fig:gelu_clipping_search} and \Cref{fig:gelu_search}. In \Cref{fig:gelu_clipping_search}, we evaluate the tradeoff between accuracy and LUT entries in the LUT domain. A larger domain covers a wider non-linear range; however, more bits are required to preserve the same resolution. A similar tradeoff is observed with the LUT entries. For a fixed LUT domain, a smaller bitwidth leads to fewer LUT entries, reducing circuit area at the cost of accuracy loss. In the figures, we show that a minimal domain, $a = 3$, and a minimal bitwidth of 4 are required for MXInt ViTs. These small values make our LUT-based approach amenable to mapping onto an area-efficient datapath.

We compare our optimization with related work in \Cref{tab:gelu_comp}.
Both Huang \textit{et al.}~\cite{huang2023integer} and HeatViT~\cite{dong2023heatvit} exploit the integer-based approximation in \Cref{eqn:gelu_approx}.
They preserve high accuracy, but the bitwidth remains large.
SDA~\cite{yang2024sda} approximates GELU into the ReLU6 function~\cite{relu6} for Stable Diffusion models but loses significant accuracy for ViTs due to their sensitivity to precision.
In comparison, our approach achieves the best accuracy at the smallest bitwidth. Our MXInt-specific optimization also saves a significant number of entries compared to the vanilla LUT-based MXInt implementation. 

\begin{table}[t]
\centering
\caption{Comparison with related work on GELU optimization for DeiT-Tiny. Our work achieves the minimal accuracy loss on fabric at the smallest bitwidth.}
\label{tab:gelu_comp}
\resizebox{\linewidth}{!}{%
    \begin{tabular}{lcrrr}
        \toprule
        Methods & On Fabric & \makecell[c]{Bitwidth} & \makecell[c]{Accuracy} & \makecell[c]{Accuracy Loss} \\
        \midrule
        Original & \xmark & 16 & 72.13\% & - \\
        Auto-ViT-Acc\cite{li2022auto} & \xmark & 16 & 72.13\% & 0 \\
        HeatViT~\cite{dong2023heatvit} & \cmark & 8 & 71.95\% & 0.137\% \\
        \cite{huang2023integer} & \cmark & 8 & 71.95\% & 0.137\%\\
        SDA~\cite{yang2024sda} & \cmark & 8 & 70.03\% & 2.132\%\\
        \midrule
        Vanilla MXInt & \cmark& 14 & 72.102\% & 0.030\% \\
        Optimized MXInt & \cmark & {\bf \color{jcgreen} 5} & 72.084\% & {\bf \color{jcgreen} 0.048\%}\\
        \bottomrule
    \end{tabular}%
}
\end{table}

\subsubsection{Softmax Approximation}

The Softmax function rescales the elements in a tensor to be in the range between 0 and 1, with the sum of the output tensor being 1~\cite{softmax}. The standard arithmetic expression of the Softmax operation is as follows.
\begin{align}
y_i = \frac{exp(x_i)}{\sum_j exp(x_j)} \label{eqn:softmax_standard}
\end{align}
$x_i$ and $y_i$ represent the $i$th elements of the input and output tensors, respectively.
This operation is expensive in circuit area because it requires hardware units to perform exponential functions and divisions. These hardware units must be implemented in a general form to compute with arbitrary values. 

Integer-based datapath optimization for Softmax has been widely studied.
A popular approximation is to replace it with the subtraction of the maximum element of the sum dimension~\cite{li2023vit, llamacpp, huang2023integer, dong2023heatvit, dong2023packqvit, yang2024sda}. However, this method cannot be directly applied to MXInt values due to their shared exponent. The {\tt llama.cpp} project~\cite{llamacpp} proposes an optimization that separates the mantissa and exponent parts of a floating-point value for efficient computation. However, this method targets the CPU architecture and still relies on its efficient $exp$ hardware peripherals.


\begin{figure}
    \centering
\begin{tikzpicture}[thick,scale=0.95, every node/.style={scale=0.95}]
\pgfplotsset{every x tick label/.append style={font=\footnotesize}, compat=1.3}
\begin{semilogyaxis}[
    height=40mm,
    width=0.5\textwidth,
    xlabel={\footnotesize Bitwidth of $LUT_{exp}$},
    ytick = {1, 10, 100},
    ylabel={\footnotesize Accuracy loss in \%},
    xmin=0, xmax=9,
    ymode=log, log ticks with fixed point, 
    legend cell align={left},
    legend style={at={(0.95,0.95)},anchor=north east, draw=none, fill=white, font=\footnotesize},
    ]

\addplot[jcblue, draw=jcblue, line width=1pt, mark=triangle*] coordinates {
    (1, 72.032)
    (2, 00.122)
    (3, 00.02)
    (4, 00.048)
    (5, 00.006)
    (6, 00.008)
    (7, 00.022)
    (8, 00.018)
};

\addplot[jcgreen, draw=jcgreen, line width=1pt, mark=diamond*] coordinates {
    (1, 79.71)
    (2, 00.162)
    (3, 00.01)
    (4, 00.038)
    (5, 00.02)
    (6, 00.004)
    (7, 00.002)
    (8, 00.016)
};

\addplot[jcred, draw=red, line width=1pt, mark=pentagon*] coordinates {
    (1, 81.73)
    (2, 00.02)
    (3, 00.03)
    (4, 00.018)
    (5, 00.042)
    (6, 00.016)
    (7, 00.036)
    (8, 00.022)
};

\addplot [dashed, black] coordinates {
    (0.005, 1) (9, 1)
};

\legend{
    DeiT-Tiny, 
    DeiT-Small, 
    DeiT-Base,
}
\end{semilogyaxis}
\end{tikzpicture}    
    \caption{Accuracy loss over different bitwidths of $LUT_{exp}$.}
    \label{fig:softmax_search}
\end{figure}
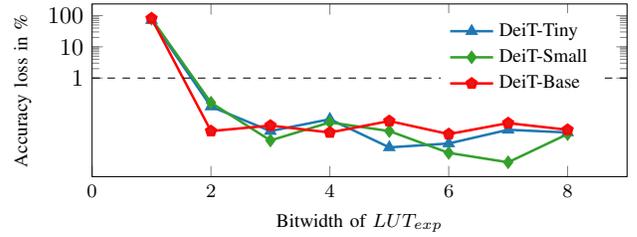

\begin{table}[t]
\centering
\caption{Comparison with related work on Softmax optimization for DeiT-Tiny. Our work achieves the minimal accuracy loss on fabric at the smallest bitwidth.}
\label{tab:softmax_comp}
\resizebox{\linewidth}{!}{%
    \begin{tabular}{lcrrr}
        \toprule
        Methods & On Fabric & \makecell[c]{Bitwidth} & \makecell[c]{Accuracy} & \makecell[c]{Accuracy Loss} \\
        \midrule
        Original & \xmark & 16 & 72.13\% & - \\
        Auto-ViT-Acc\cite{li2022auto} & \xmark & 16 & 72.13\% & 0 \\
        HeatViT~\cite{dong2023heatvit} & \cmark & 8 & 71.65\% & 0.482\% \\
        \cite{huang2023integer} & \cmark & 8 & 71.916\% & 0.216\%\\
        SDA~\cite{yang2024sda} & \cmark & 8 & 72.12\% & 0.012\%\\
        \midrule
        Vanilla MXInt & \cmark& 16 & 72.148\% & +0.016\% \\
        MXInt to match SDA & \cmark & {5} & 72.118\% & 0.014\%\\
        Optimized MXInt & \cmark & {\bf \color{jcgreen} 2} & 72.01\% & {\bf \color{jcgreen} 0.122\%}\\
        \bottomrule
    \end{tabular}%
}
\end{table}

We adopt the approach by {\tt llama.cpp} and extend it to MXInt datapath optimization.
Specifically, we propose a LUT-based approach that transforms the exponent $e^x$ into an expression of integers $n$ and a fixed-point number $r$ as follows.
\begin{align}
e^x &= 2^{x\log_2^e} \\
n   &= \lfloor fp2Int (x\log_2(e)) \rfloor \\
r   &= x\log_2(e) - n \\
e^x &= 2^{n + r} = 2^n \cdot 2^r
\label{eqn:approx_exp_old}
\end{align}
$fp2Int$ casts the floating-point format to fixed-point numbers.
Instead of computing $2^r$ using arithmetic operators, we choose to map the function into a LUT, shown as follows. 
\begin{align}
2^r &\approx LUT_{pow2}(r) \\
e^x &\approx 2^n \cdot LUT_{pow2}(r)
\label{eqn:approx_exp_new}
\end{align}

Benefiting from the efficient representation of MXInt, only a small number of bits are required to represent $n$ and $r$. $n$ is handled by the shift operation, and its shift bits are determined by the bitwidth of $r$. An efficient $r$ could significantly save both the unrolled shift operator area and the number of LUT entries.

A key benefit of this optimization is that the output values are already represented in a floating-point number format ($2^n \cdot LUT_{pow2}(r) = 2^{x_e} x_m$), and can be passed to perform a division efficiently, as shown below:
\begin{align}
\frac{x_1}{x_2} = \frac{x_{m1} 2^{x_{e1}}}{x_{m2} 2^{x_{e2}}} = \frac{x_{m1}}{x_{m2}} 2^{x_{e1} - x_{e2}}
\end{align}
This leads to a significant reduction in circuit area compared to the vanilla approach that casts all values to floating-point before computing $exp$. 

\Cref{fig:softmax_search} shows the design space exploration for the bitwidth of $r$ over different models. We observed that the bitwidth required for $r$ is small, only requiring two bits to preserve high model accuracy. This leads to an efficient datapath design for the exponential function in MXInt. \Cref{tab:softmax_comp} compares our design with related work. We present an intermediate design point to compare with the state-of-the-art design by SDA~\cite{yang2024sda} and show that our approach achieves the highest accuracy with the smallest bitwidth. The optimized datapath in our final designs only requires two bits under the accuracy loss budget.

\begin{table}[]
    \centering
    \caption{Model accuracy over different quantization techniques and precisions. MXInt quantization achieves lower bits compared to traditional approaches. MXInt$N$ means $N$-bit mantissa. The exponent is always 8-bit in MXInt. The block size is 16 for activations and 256 for parameters.}
    \label{tab:quantization}
\resizebox{0.48\textwidth}{!}{%
\begin{tabular}{ccrrrrrr}
\toprule
\multicolumn{2}{c}{Precision} & \multicolumn{2}{c}{DeiT Tiny } & \multicolumn{2}{c}{DeiT Small } & \multicolumn{2}{c}{DeiT Base } \\
\cmidrule(lr){1-2}
\cmidrule(lr){3-4}
\cmidrule(lr){5-6}
\cmidrule(lr){7-8}
Parameters & Activations & \multicolumn{1}{c}{\%} & \multicolumn{1}{c}{$\Delta$\%} & \multicolumn{1}{c}{\%} & \multicolumn{1}{c}{$\Delta$\%} & \multicolumn{1}{c}{\%} & \multicolumn{1}{c}{$\Delta$\%} \\
\midrule
Float32 & Float32 & 72.13 & 0.00 & 79.83 & 0.00 & 81.80 & 0.00 \\
Float8 & Float8 & 71.26 & -0.87 & 79.26 & -0.57 & 81.74 & -0.06 \\
Int16 & Int16 & 71.85 & -0.28 & 79.34 & -0.49 & 80.05 & -1.75 \\
Int8 & Int8 & 0.10 & \textcolor{jcred}{-72.03} & 0.11 & \textcolor{jcred}{-79.72} & 0.10 & \textcolor{jcred}{-81.70} \\
MXInt8 & MXInt8 & 72.04 & -0.09 & 79.80 & -0.03 & 81.84 & 0.04 \\
MXInt6 & MXInt8 & 71.56 & -0.57 & 79.42 & -0.41 & 81.72 & -0.08 \\
MXInt6 & MXInt6 & 70.97 & \textcolor{jcred}{-1.16 } & 79.13 & -0.70 & 81.70 & -0.10 \\
MXInt4 & MXInt6 & 54.61 & \textcolor{jcred}{-17.52} & 70.21 & \textcolor{jcred}{-9.62} & 77.53 & \textcolor{jcred}{-4.27} \\
\bottomrule
\end{tabular} 

}
\end{table}

\section{Experiments}
\label{sec:experiments}


In this section, we evaluate the proposed MXInt datapath optimization by addressing the following questions. 
\begin{enumerate}[label={\arabic*)}]
    \item What are the bitwidth saving breakdown by quantization the whole ViTs in MXInt? 
    \item What are the system speedup achieved by MXInt datapath optimization bring? 
    \item What are the insights for designing future MXInt accelerators?
\end{enumerate}

We evaluated our work on the DeiT family~\cite{touvron2022deit}, including DeiT Tiny, DeiT Small, and DeiT Base, on the ImageNet dataset~\cite{deng2009imagenet}. 
All of them were obtained directly from PyTorch Image Models~\cite{timm}. 
We evaluated the model accuracy on PTQ results to show the scalability of our approach so that no GPU computing time is required for fine-tuning. 
We target Alveo U250 FPGAs due to their availability, but our datapath optimization results are independent of the FPGA platform. The version of Xilinx software used was 2023.2. 
Our results were obtained from cycle-accurate simulation and the implementation reports from Vivado. 
The results of related work were obtained from their publications. To ensure fairness, we compare our work with the non-pruned HeatViT~\cite{dong2023heatvit} and focus on MXInt datapath optimization.

\begin{table}[t]
\caption{Area saving in LUT entries and accuracy loss by our approximation techniques.
The LUT entries are reduced by at least 16$\times$.
The bitwidths are determined by greedy search.}
\label{tab:approx_ops}
\resizebox{0.48\textwidth}{!}{%
\begin{tabular}{llrrrr}
\toprule
\multicolumn{1}{c}{\multirow{2}{*}{Operation}} & \multicolumn{1}{c}{\multirow{2}{*}{Approach}} & \multicolumn{1}{c}{\multirow{2}{*}{\makecell[c]{LUT \\ entry bits}}} & \multicolumn{3}{c}{Accuracy loss in \%} \\
\cmidrule(lr){4-6}
\multicolumn{1}{c}{} & \multicolumn{1}{c}{} & \multicolumn{1}{c}{} & \multicolumn{1}{c}{DeiT Tiny} & \multicolumn{1}{c}{DeiT Small} & \multicolumn{1}{c}{DeiT Base} \\
\midrule
 & Float32 & - & 79.83 & 72.13 & 81.80 \\
\midrule
\multirow{2}{*}{GELU} & vanilla LUT & 14 & 0.03 & 0.04 & 0.01 \\
 & Our work & \textbf{\color{jcgreen} 5} & 0 & 0.07 & 0.06 \\
\midrule
\multirow{2}{*}{Softmax} & vanilla LUT & 16 & 0.01 & 0 & 0.03 \\
 & Our work & \textbf{\color{jcgreen} 2} & 0.03 & 0.02 & 0.03 \\
\midrule
\multirow{2}{*}{LayerNorm} & vanilla LUT & 13 & 0.11 & 0.14 & 0.06 \\
 & Our work & \textbf{\color{jcgreen} 5} & 0.55 & 0.43 & 0.08 \\
\bottomrule
\end{tabular}
}
\end{table}
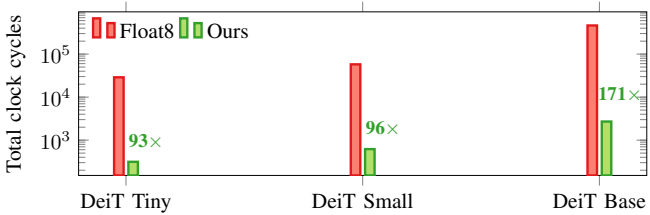
\begin{figure}[t]
\begin{tikzpicture}[
thick,scale=0.9, every node/.style={scale=0.9},
]
\pgfplotsset{compat=1.3}
\begin{axis}[
width=0.55\textwidth,
legend columns = 4,
height=4cm,
ybar, 
enlarge x limits={abs=0.7cm},
bar width=4,
ymode=log, 
symbolic x coords={
tiny,
small,
base,
},
ylabel={Total clock cycles},
xtick=data,
xticklabels={
DeiT~Tiny,
DeiT~Small,
DeiT~Base,
},
legend style={at={(0.01,0.88)},anchor=west, draw=none, fill=none}
]

\addplot[jcredl, draw=jcred, fill=jcredl, line width=1pt] table
[x=x, y=fp32, col sep=space] {
x	fp32	b1	ours
tiny	28901	931	314
small	57802	1806	1207
base	462422	2709	2709
};


\addplot[
jcgreenl, draw=jcgreen, fill=jcgreenl, line width=1pt,
] table  
[x=x, y=ours, col sep=space, meta=speedup] {
x	fp32	b1	ours	speedup
tiny	28901	931	314	92$\times$
small	115605	1806	620	95$\times$
base	2774532	2709	2709	1024$\times$
};

\legend{Float8, Ours}

\end{axis}

\node[draw=none, align=left, font=\footnotesize, color=jcgreen] at (1.0, 0.5) {\bf 93$\times$};
\node[draw=none, align=left, font=\footnotesize, color=jcgreen] at (4.5, 0.7) {\bf 96$\times$};
\node[draw=none, align=left, font=\footnotesize, color=jcgreen] at (8, 1.2) {\bf 171$\times$};
\end{tikzpicture}
\caption{
Speedups achieved by our work (shown in \textcolor{jcgreen}{green}).
}
\label{fig:speedup}
\end{figure}
\begin{table*}[]
    \centering
    \caption{Comparing our work with related work on ViT accelerators is unfair due to the differences in hardware microarchitectures (e.g. systolic arrays and dataflow circuits) and algorithm optimizations (e.g. fine-tuning and PTQ). In this work, we focus on MXInt datapath optimization and only report our system results of MXInt ViTs to be self-contained.}
    \label{tab:main_results}
\resizebox{\textwidth}{!}{%
\begin{tabular}{lllrrrrrrrrrrrrr}
\toprule
\multicolumn{1}{c}{\multirow{2}{*}{Models}} & \multicolumn{1}{c}{\multirow{2}{*}{Methods}} & \multicolumn{1}{c}{\multirow{2}{*}{Platforms}} & \multicolumn{9}{c}{Hardware resources} & \multicolumn{4}{c}{Performance} \\
\cmidrule(lr){4-11}
\cmidrule(lr){12-16}
\multicolumn{1}{c}{} & \multicolumn{1}{c}{} & \multicolumn{1}{c}{} & \multicolumn{1}{c}{kLUTs} & \multicolumn{1}{c}{\%} & \multicolumn{1}{c}{DSPs} & \multicolumn{1}{c}{\%} & \multicolumn{1}{c}{BRAM} & \multicolumn{1}{c}{\%} & \multicolumn{1}{c}{URAM} & \multicolumn{1}{c}{\%} & \multicolumn{1}{c}{Power (W)} & \multicolumn{1}{c}{Fmax (MHz)} & \multicolumn{1}{c}{GOPs/s} & \multicolumn{1}{c}{FPS} & \multicolumn{1}{c}{FPS/W} \\
\midrule
\multirow{3}{*}{DeiT Tiny} & \cite{huang2023integer} & ZCU102 & 144 & 52.7 & 1268 & 50.3 & - & - & - & - & - & 300 & 616.10 & 245.28 & - \\
 & HeatViT~\cite{dong2023heatvit} & ZCU102 & 116 & 42 & 1739 & 69 & 289 & 32 & - & - & 8.01 & 150 & 197.86 & 78.30 & 9.77 \\
 & Ours & U250 & 1163 & 67 & 24 & 0.2 & 330 & 12 & 529 & 41 & 40.08 & 183 & 1488.34 & 589.44 & 14.71 \\
\midrule
\multirow{4}{*}{DeiT Small} & \cite{huang2023integer} & ZCU102 & 144 & 52.7 & 1268 & 50.3 & - & - & - & - & - & 300 & 762.70 & 89.69 & - \\
 & HeatViT~\cite{dong2023heatvit} & ZCU102 & 130 & 48 & 1754 & 70 & 493 & 54 & - & - & 10.10 & 150 & 239.81 & 25.90 & 2.57 \\
 & Auto-ViT-Acc~\cite{li2022auto} & ZCU102 & 185 & 67 & 1552 & 62 & - & - & - & - & 9.63 & 150 & 907.80 & 99.70 & 10.35 \\
 & Ours & U250 & 860 & 50 & 12 & 0.1 & 1476 & 55 & 820 & 64 & 28.37 & 192 & 2861.03 & 309.21 & 10.89 \\
 \midrule
\multirow{3}{*}{DeiT Base} & HeatViT~\cite{dong2023heatvit} & ZCU102 & 145 & 53 & 1786 & 71 & 664 & 73 & - & - & 11.04 & 150 & 395.80 & 11.20 & 1.01 \\
 & Auto-ViT-Acc~\cite{li2022auto} & ZCU102 & 186 & 68 & 1556 & 62 & - & - & - & - & 9.31 & 150 & 1181.50 & 34.00 & 3.65 \\
 & Ours & U250 & 744 & 43 & 8 & 0.1 & 497 & 18 & 1089 & 85 & 25.25 & 179 & 2332.37 & 66.06 & 2.61 \\
 \bottomrule
\end{tabular}}
\end{table*}

\subsubsection{Bitwidth Saving Evaluation}
\label{sec:quant_results}

We first discuss the effectiveness of our work based on model accuracy.
Table~\ref{tab:quantization} shows the model accuracy over a set of quantization approaches.
As mentioned before, we focus on PTQ results to ensure fairness, as different approaches may have their own fine-tuning optimizations. 
In the table, integer quantization struggles to preserve high accuracy with smaller bits. 
This is because the numerical variance between tensor elements is huge and requires a large range to be represented.
On the other hand, both the standard floating-point format and the MXInt format have an exponent to scale these values, leading to a large dynamic range. 
With the same bitwidth, MXInt shows better accuracy compared to Float8 because it has five more mantissa bits at the price of sharing the exponent.
We apply a greedy quantization search and determine the minimal bitwidth required for lossless quantization in each ViT model, as shown in the table.
The final precision is used for hardware mapping in Table~\ref{tab:main_results}.
Finally, it can also be seen that larger models tend to have better tolerance to low-bit quantization, which suggests our approach will be more effective as we expect future model sizes will continue to increase.
Also, the PTQ results avoid GPU time on retraining, making our quantization approach more applicable when training resources are limited.

The evaluation of our datapath optimization on non-linear operations at the system level is shown in Table~\ref{tab:approx_ops}.
In the table, it can be seen that a significant number of entries are saved by our datapath optimizations, where reducing a bit leads to half of the LUT size. 
Besides high area efficiency, our approximation techniques preserve high model accuracy, and the accuracy loss is negligible.

\subsubsection{System Performance Evaluation}
\label{sec:speedup_results}

We also compared our work with the corresponding floating-point implementation \Cref{fig:speedup}.
The red bars represent the same architecture in Float8, the green bars represent our work without and with inter-layer pipelining, respectively.
We chose Float8 rather than Float32 as our baseline because the 8-bit floating-point format has a similar bitwidth to our MXInt format, leading to a fair comparison.
We compare our work with the same architecture in Float8 to show the effectiveness of MXInt mapping. 
The independent exponent operators in Float8 lead to significant overhead in both area and memory size, which limits the design space for data parallelism and inter-layer pipelining.
The latency is significantly worse than that of the MXInt design in the green bars.
Our approach achieves at least a 93$\times$ speedup by translating Float8 operators into efficient MXInt hardware in a lossless form.
Still, the speedup for large models is restricted by the available hardware resources on the FPGA and can be further exploited in larger FPGAs such as the V80 device~\cite{v80}.

\subsubsection{Discussion on Limitations}
\label{sec:related_work_results}

We include related work on ViT accelerators in Table~\ref{tab:main_results} with our work.
The related work cannot be directly compared to our results due to two main reasons. First, they map part of their models on the CPU, while we map the whole models on the fabric, leading to different energy efficiencies.
Second, they fine-tune the model using different software optimization techniques, while we focus on datapath optimization without fine-tuning.
Our quantization results offer a lower bound for our approach and can be significantly improved with fine-tuning.

In the table, existing work focuses on fixed-point quantization and acceleration.
Their fixed-point operators fully exploit existing DSP blocks for acceleration, as the DSPs are hardened on-chip and amenable to efficient fixed-point/floating-point multiplication and addition.
In our work, the DSP blocks are significantly underutilized as we focus on general datapath optimization for both ASIC and FPGA designs.
The proposed accelerator is bounded by the LUT resources as most of the operators are mapped into LUTs.
Efficient logic synthesis for MXInt operators to FPGA-specific IPs, such as DSP blocks, is outside our scope but is key future work, where our work provides an initial baseline.

\section{Related Work}
\label{sec:related_work}

{\em Microscaling Quantization:}
Sharing certain components for a block of values has been widely recognized as the state-of-the-art technique for quantizing CNNs~\cite{lin2017accurate, zhang2018lqnets}. Further explorations within this line of research have investigated grouping numbers at various granularities, including layer-wise~\cite{wu2018training}, channel-wise~\cite{krishnamoorthi2018quantizing}, and vector-wise quantization~\cite{dai2021vs}. In addition, many block floating-point variants~\cite{harma2022accuracy, dai2021vs, darvish2020pushing} have been proposed, with the core idea of grouping values into multiple blocks and elements within each block sharing common digits.
The closest piece related to our work is by Darvish \textit{et al.}~\cite{darvish2020pushing} that proposes MXInt quantization for DNN accelerators. This work is later extended to multi-level MX quantization, also known as MXFP, where the shared component can be non-integers~\cite{darvish2023shared}. They focus on MXInt quantization and overlook hardware optimization, while our work proposes MXInt-specific datapath optimization with design space exploration.

{\em Quantized Transformer Accelerators:}
Quantization for efficient ML inference on accelerators has been widely studied~\cite{andri2022going, song2020drq, zadeh2022mokey, zhao2021cambricon}, especially using fixed-point numbers~\cite{dettmers2022llm, frantar2022gptq, xiao2022smoothquant, yao2022zeroquant, liu2023psq, liu2024spark}. Other work customizes hardware architectures for efficient inference~\cite{wu2023msd, fan2022adaptable, ham20203, hong2022dfx, kao2023flat, lu2021sanger}. GOBO~\cite{zadeh2020gobo} and EdgeBERT~\cite{tambe2021edgebert} exploit software and hardware co-designs for accelerating transformers. FACT~\cite{qin2023fact} and FlightLLM~\cite{zeng2024flightllm} exploit mixed-precision quantization using fixed-point numbers on linear layers. They only exploit quantization with fixed-point numbers, while we target MXInt quantization.

In the  domain of ViT accelerators, existing work focuses on fixed-point quantization~\cite{li2022auto, dong2023heatvit, huang2023integer}, while we propose MXInt quantization with hardware optimizations. They only accelerate part of the models on the FPGA, while our hardware accelerator computes the complete workload of the model.

\section{Conclusions}
\label{sec:conclusion}

In this work, we propose fully hardware-accelerated ViTs in MXInt and customize hardware datapath optimization for MXInt operators.
We explore the tradeoff between model accuracy and area efficiency of the MXInt hardware operators and determine a balance for efficient acceleration.
Given the block sharing property of the MXInt format, our proposed datapath optimization techniques further tailor operators with complex circuitry to efficient LUTs.
Our optimization realizes all operations mapped to efficient hardware on the fabric and has shown significant improvements compared to vanilla floating-point designs.

\section*{Acknowledgments}

This work was supported by the Advanced Research + Invention Agency (ARIA), UK. We also thank ARIA for their research network.


\newpage
\balance
\bibliographystyle{ieeetr}
\bibliography{refs}

\begin{thebibliography}{10}

\bibitem{zeng2024flightllm}
S.~Zeng, J.~Liu, G.~Dai, X.~Yang, T.~Fu, H.~Wang, W.~Ma, H.~Sun, S.~Li, Z.~Huang, {\em et~al.}, ``Flightllm: Efficient large language model inference with a complete mapping flow on fpga,'' {\em arXiv preprint arXiv:2401.03868}, 2024.

\bibitem{dong2023heatvit}
P.~Dong, M.~Sun, A.~Lu, Y.~Xie, K.~Liu, Z.~Kong, X.~Meng, Z.~Li, X.~Lin, Z.~Fang, {\em et~al.}, ``Heatvit: Hardware-efficient adaptive token pruning for vision transformers,'' in {\em 2023 IEEE International Symposium on High-Performance Computer Architecture (HPCA)}, pp.~442--455, IEEE, 2023.

\bibitem{li2022auto}
Z.~Li, M.~Sun, A.~Lu, H.~Ma, G.~Yuan, Y.~Xie, H.~Tang, Y.~Li, M.~Leeser, Z.~Wang, {\em et~al.}, ``Auto-vit-acc: An fpga-aware automatic acceleration framework for vision transformer with mixed-scheme quantization,'' in {\em 2022 32nd International Conference on Field-Programmable Logic and Applications (FPL)}, pp.~109--116, IEEE, 2022.

\bibitem{alexey2020image}
D.~Alexey, ``An image is worth 16x16 words: Transformers for image recognition at scale,'' {\em arXiv preprint arXiv: 2010.11929}, 2020.

\bibitem{yang2024sda}
G.~Yang, Y.~Xie, Z.~J. Xue, S.-E. Chang, Y.~Li, P.~Dong, J.~Lei, W.~Xie, Y.~Wang, X.~Lin, {\em et~al.}, ``Sda: Low-bit stable diffusion acceleration on edge fpgas,'' in {\em 2024 34th International Conference on Field-Programmable Logic and Applications (FPL)}, pp.~264--273, IEEE, 2024.

\bibitem{darvish2020pushing}
B.~Darvish~Rouhani, D.~Lo, R.~Zhao, M.~Liu, J.~Fowers, K.~Ovtcharov, A.~Vinogradsky, S.~Massengill, L.~Yang, R.~Bittner, {\em et~al.}, ``Pushing the limits of narrow precision inferencing at cloud scale with microsoft floating point,'' {\em Advances in neural information processing systems}, vol.~33, pp.~10271--10281, 2020.

\bibitem{zhang2023revisiting}
C.~Zhang, J.~Cheng, I.~Shumailov, G.~A. Constantinides, and Y.~Zhao, ``Revisiting block-based quantisation: What is important for sub-8-bit llm inference?,'' 2023.

\bibitem{samson2024exploring}
E.~Samson, N.~Mellempudi, W.~Luk, and G.~A. Constantinides, ``Exploring fpga designs for mx and beyond,'' in {\em 2024 34th International Conference on Field-Programmable Logic and Applications (FPL)}, pp.~304--310, IEEE, 2024.

\bibitem{huang2023integer}
M.~Huang, J.~Luo, C.~Ding, Z.~Wei, S.~Huang, and H.~Yu, ``An integer-only and group-vector systolic accelerator for efficiently mapping vision transformer on edge,'' {\em IEEE Transactions on Circuits and Systems I: Regular Papers}, 2023.

\bibitem{touvron2022deit}
H.~Touvron, M.~Cord, and H.~J{\'e}gou, ``Deit iii: Revenge of the vit,'' in {\em European conference on computer vision}, pp.~516--533, Springer, 2022.

\bibitem{deng2009imagenet}
J.~Deng, W.~Dong, R.~Socher, L.-J. Li, K.~Li, and L.~Fei-Fei, ``Imagenet: A large-scale hierarchical image database,'' in {\em 2009 IEEE conference on computer vision and pattern recognition}, pp.~248--255, Ieee, 2009.

\bibitem{kahan1996ieee}
W.~Kahan, ``Ieee standard 754 for binary floating-point arithmetic,'' {\em Lecture Notes on the Status of IEEE}, vol.~754, no.~94720-1776, p.~11, 1996.

\bibitem{chen2024understanding}
H.~Chen, J.~Zhang, Y.~Du, S.~Xiang, Z.~Yue, N.~Zhang, Y.~Cai, and Z.~Zhang, ``Understanding the potential of fpga-based spatial acceleration for large language model inference,'' {\em ACM Transactions on Reconfigurable Technology and Systems}, 2024.

\bibitem{venieris2016fpgaconvnet}
S.~I. Venieris and C.-S. Bouganis, ``fpgaconvnet: A framework for mapping convolutional neural networks on fpgas,'' in {\em 2016 IEEE 24th Annual International Symposium on Field-Programmable Custom Computing Machines (FCCM)}, pp.~40--47, IEEE, 2016.

\bibitem{umuroglu2017finn}
Y.~Umuroglu, N.~J. Fraser, G.~Gambardella, M.~Blott, P.~Leong, M.~Jahre, and K.~Vissers, ``Finn: A framework for fast, scalable binarized neural network inference,'' in {\em Proceedings of the 2017 ACM/SIGDA international symposium on field-programmable gate arrays}, pp.~65--74, 2017.

\bibitem{ye2023hida}
H.~Ye, H.~Jun, and D.~Chen, ``Hida: A hierarchical dataflow compiler for high-level synthesis,'' {\em arXiv preprint arXiv:2311.03379}, 2023.

\bibitem{coward2023automating}
S.~Coward, G.~A. Constantinides, and T.~Drane, ``Automating constraint-aware datapath optimization using e-graphs,'' in {\em 2023 60th ACM/IEEE Design Automation Conference (DAC)}, pp.~1--6, IEEE, 2023.

\bibitem{alexandridis2024online}
K.~Alexandridis and G.~Dimitrakopoulos, ``Online alignment and addition in multiterm floating-point adders,'' {\em IEEE Transactions on Very Large Scale Integration (VLSI) Systems}, 2024.

\bibitem{layernorm}
{PyTorch LayerNorm}, 2025.

\bibitem{gelu}
{PyTorch GELU}, 2025.

\bibitem{relu6}
{PyTorch ReLu6}, 2025.

\bibitem{softmax}
{PyTorch Softmax}, 2025.

\bibitem{li2023vit}
Z.~Li and Q.~Gu, ``I-vit: Integer-only quantization for efficient vision transformer inference,'' in {\em Proceedings of the IEEE/CVF International Conference on Computer Vision}, pp.~17065--17075, 2023.

\bibitem{llamacpp}
{llama.cpp}, 2025.

\bibitem{dong2023packqvit}
P.~Dong, L.~Lu, C.~Wu, C.~Lyu, G.~Yuan, H.~Tang, and Y.~Wang, ``Packqvit: Faster sub-8-bit vision transformers via full and packed quantization on the mobile,'' {\em Advances in Neural Information Processing Systems}, vol.~36, pp.~9015--9028, 2023.

\bibitem{timm}
{Pytorch Image Models}, 2023.

\bibitem{v80}
{AMD Alveo V80 Compute Accelerator}, 2025.

\bibitem{lin2017accurate}
X.~Lin, C.~Zhao, and W.~Pan, ``Towards accurate binary convolutional neural network,'' 2017.

\bibitem{zhang2018lqnets}
D.~Zhang, J.~Yang, D.~Ye, and G.~Hua, ``Lq-nets: Learned quantization for highly accurate and compact deep neural networks,'' 2018.

\bibitem{wu2018training}
S.~Wu, G.~Li, F.~Chen, and L.~Shi, ``Training and inference with integers in deep neural networks,'' {\em arXiv preprint arXiv:1802.04680}, 2018.

\bibitem{krishnamoorthi2018quantizing}
R.~Krishnamoorthi, ``Quantizing deep convolutional networks for efficient inference: A whitepaper,'' {\em arXiv preprint arXiv:1806.08342}, 2018.

\bibitem{dai2021vs}
S.~Dai, R.~Venkatesan, M.~Ren, B.~Zimmer, W.~Dally, and B.~Khailany, ``Vs-quant: Per-vector scaled quantization for accurate low-precision neural network inference,'' {\em Proceedings of Machine Learning and Systems}, vol.~3, pp.~873--884, 2021.

\bibitem{harma2022accuracy}
S.~B. Harma, C.~S{\"o}nmez, B.~Falsafi, M.~Jaggi, and Y.~Oh, ``Accuracy boosters: Epoch-driven mixed-mantissa block floating-point for dnn training,'' {\em arXiv preprint arXiv:2211.10737}, 2022.

\bibitem{darvish2023shared}
B.~Darvish~Rouhani, R.~Zhao, V.~Elango, R.~Shafipour, M.~Hall, M.~Mesmakhosroshahi, A.~More, L.~Melnick, M.~Golub, G.~Varatkar, {\em et~al.}, ``With shared microexponents, a little shifting goes a long way,'' in {\em Proceedings of the 50th Annual International Symposium on Computer Architecture}, pp.~1--13, 2023.

\bibitem{andri2022going}
R.~Andri, B.~Bussolino, A.~Cipolletta, L.~Cavigelli, and Z.~Wang, ``Going further with winograd convolutions: Tap-wise quantization for efficient inference on 4x4 tiles,'' in {\em 2022 55th IEEE/ACM International Symposium on Microarchitecture (MICRO)}, pp.~582--598, IEEE, 2022.

\bibitem{song2020drq}
Z.~Song, B.~Fu, F.~Wu, Z.~Jiang, L.~Jiang, N.~Jing, and X.~Liang, ``Drq: dynamic region-based quantization for deep neural network acceleration,'' in {\em 2020 ACM/IEEE 47th Annual International Symposium on Computer Architecture (ISCA)}, pp.~1010--1021, IEEE, 2020.

\bibitem{zadeh2022mokey}
A.~H. Zadeh, M.~Mahmoud, A.~Abdelhadi, and A.~Moshovos, ``Mokey: Enabling narrow fixed-point inference for out-of-the-box floating-point transformer models,'' in {\em Proceedings of the 49th Annual International Symposium on Computer Architecture}, pp.~888--901, 2022.

\bibitem{zhao2021cambricon}
Y.~Zhao, C.~Liu, Z.~Du, Q.~Guo, X.~Hu, Y.~Zhuang, Z.~Zhang, X.~Song, W.~Li, X.~Zhang, {\em et~al.}, ``Cambricon-q: A hybrid architecture for efficient training,'' in {\em 2021 ACM/IEEE 48th Annual International Symposium on Computer Architecture (ISCA)}, pp.~706--719, IEEE, 2021.

\bibitem{dettmers2022llm}
T.~Dettmers, M.~Lewis, Y.~Belkada, and L.~Zettlemoyer, ``Llm. int8 (): 8-bit matrix multiplication for transformers at scale,'' {\em arXiv preprint arXiv:2208.07339}, 2022.

\bibitem{frantar2022gptq}
E.~Frantar, S.~Ashkboos, T.~Hoefler, and D.~Alistarh, ``Gptq: Accurate post-training quantization for generative pre-trained transformers,'' {\em arXiv preprint arXiv:2210.17323}, 2022.

\bibitem{xiao2022smoothquant}
G.~Xiao, J.~Lin, M.~Seznec, J.~Demouth, and S.~Han, ``Smoothquant: Accurate and efficient post-training quantization for large language models,'' {\em arXiv preprint arXiv:2211.10438}, 2022.

\bibitem{yao2022zeroquant}
Z.~Yao, R.~Yazdani~Aminabadi, M.~Zhang, X.~Wu, C.~Li, and Y.~He, ``Zeroquant: Efficient and affordable post-training quantization for large-scale transformers,'' {\em Advances in Neural Information Processing Systems}, vol.~35, pp.~27168--27183, 2022.

\bibitem{liu2023psq}
F.~Liu, N.~Yang, and L.~Jiang, ``Psq: An automatic search framework for data-free quantization on pim-based architecture,'' in {\em 2023 IEEE 41st International Conference on Computer Design (ICCD)}, pp.~507--514, IEEE, 2023.

\bibitem{liu2024spark}
F.~Liu, N.~Yang, H.~Li, Z.~Wang, Z.~Song, S.~Pei, and L.~Jiang, ``Spark: Scalable and precision-aware acceleration of neural networks via efficient encoding,'' in {\em 2024 IEEE International Symposium on High-Performance Computer Architecture (HPCA)}, pp.~1029--1042, IEEE, 2024.

\bibitem{wu2023msd}
J.~Wu, J.~Zhou, Y.~Gao, Y.~Ding, N.~Wong, and H.~K.-H. So, ``Msd: Mixing signed digit representations for hardware-efficient dnn acceleration on fpga with heterogeneous resources,'' in {\em 2023 IEEE 31st Annual International Symposium on Field-Programmable Custom Computing Machines (FCCM)}, pp.~94--104, IEEE, 2023.

\bibitem{fan2022adaptable}
H.~Fan, T.~Chau, S.~I. Venieris, R.~Lee, A.~Kouris, W.~Luk, N.~D. Lane, and M.~S. Abdelfattah, ``Adaptable butterfly accelerator for attention-based nns via hardware and algorithm co-design,'' in {\em 2022 55th IEEE/ACM International Symposium on Microarchitecture (MICRO)}, pp.~599--615, IEEE, 2022.

\bibitem{ham20203}
T.~J. Ham, S.~J. Jung, S.~Kim, Y.~H. Oh, Y.~Park, Y.~Song, J.-H. Park, S.~Lee, K.~Park, J.~W. Lee, {\em et~al.}, ``A\^{} 3: Accelerating attention mechanisms in neural networks with approximation,'' in {\em 2020 IEEE International Symposium on High Performance Computer Architecture (HPCA)}, pp.~328--341, IEEE, 2020.

\bibitem{hong2022dfx}
S.~Hong, S.~Moon, J.~Kim, S.~Lee, M.~Kim, D.~Lee, and J.-Y. Kim, ``Dfx: A low-latency multi-fpga appliance for accelerating transformer-based text generation,'' in {\em 2022 55th IEEE/ACM International Symposium on Microarchitecture (MICRO)}, pp.~616--630, IEEE, 2022.

\bibitem{kao2023flat}
S.-C. Kao, S.~Subramanian, G.~Agrawal, A.~Yazdanbakhsh, and T.~Krishna, ``Flat: An optimized dataflow for mitigating attention bottlenecks,'' in {\em Proceedings of the 28th ACM International Conference on Architectural Support for Programming Languages and Operating Systems, Volume 2}, pp.~295--310, 2023.

\bibitem{lu2021sanger}
L.~Lu, Y.~Jin, H.~Bi, Z.~Luo, P.~Li, T.~Wang, and Y.~Liang, ``Sanger: A co-design framework for enabling sparse attention using reconfigurable architecture,'' in {\em MICRO-54: 54th Annual IEEE/ACM International Symposium on Microarchitecture}, pp.~977--991, 2021.

\bibitem{zadeh2020gobo}
A.~H. Zadeh, I.~Edo, O.~M. Awad, and A.~Moshovos, ``Gobo: Quantizing attention-based nlp models for low latency and energy efficient inference,'' in {\em 2020 53rd Annual IEEE/ACM International Symposium on Microarchitecture (MICRO)}, pp.~811--824, IEEE, 2020.

\bibitem{tambe2021edgebert}
T.~Tambe, C.~Hooper, L.~Pentecost, T.~Jia, E.-Y. Yang, M.~Donato, V.~Sanh, P.~Whatmough, A.~M. Rush, D.~Brooks, and G.-Y. Wei, ``Edgebert: Sentence-level energy optimizations for latency-aware multi-task nlp inference,'' in {\em MICRO-54: 54th Annual IEEE/ACM International Symposium on Microarchitecture}, MICRO '21, (New York, NY, USA), p.~830–844, Association for Computing Machinery, 2021.

\bibitem{qin2023fact}
Y.~Qin, Y.~Wang, D.~Deng, Z.~Zhao, X.~Yang, L.~Liu, S.~Wei, Y.~Hu, and S.~Yin, ``Fact: Ffn-attention co-optimized transformer architecture with eager correlation prediction,'' in {\em Proceedings of the 50th Annual International Symposium on Computer Architecture}, pp.~1--14, 2023.

\end{thebibliography}

\end{document}